\documentclass[preprint,aps,tightenlines,showpacs,nofootinbib]{revtex4}
\usepackage{latexsym}
\usepackage[dvips]{graphicx}
\newcommand{\beq}{\begin{eqnarray}}
\newcommand{\eeq}{\end{eqnarray}}
\newcommand{\eq}{eqnarray}

\newcommand{\al}{{\alpha}}
\newcommand{\be}{{\beta}}
\newcommand{\ci}{\cite}

\newcommand{\la}{{\lambda}}
\newcommand{\La}{{\Lambda}}
\newcommand{\m}{{\mu}}

\newcommand{\om}{{\omega}}
\newcommand{\Om}{{\Omega}}

\newcommand{\no}{{\nonumber}}
\newcommand{\f}{\frac}
\newcommand{\ra}{\rightarrow}

\newcommand{\Sch}{Schwarzschild }

\newcommand{\asy}{asymptotically}

\begin{document}

\preprint{arXiv:0905.4480v5
 [hep-th]}

\title{The Black Hole and Cosmological Solutions in IR modified Ho\v{r}ava Gravity}

\author{Mu-In Park\footnote{E-mail address: muinpark@gmail.com}}

\affiliation{ Research Institute of Physics and Chemistry, Chonbuk
National University, Chonju 561-756, Korea }

\begin{abstract}
Recently Ho\v{r}ava proposed a renormalizable gravity theory in four
dimensions which reduces to Einstein gravity with a {\it
non-vanishing} cosmological constant in IR but with improved UV
behaviors. Here, I study an IR modification
which breaks ``softly" the detailed balance condition in Ho\v{r}ava
model and allows the asymptotically {\it flat} limit as well. I
obtain the black hole and cosmological solutions for ``arbitrary"
cosmological constant that represent the analogs of the standard
Schwarzschild-(A)dS solutions which can be asymptotically (A)dS as
well as flat and I discuss their thermodynamical properties. I also
obtain solutions for FRW metric with an arbitrary cosmological
constant. I study its implication to the dark energy and find that
it seems to be consistent with current observational data.

\end{abstract}


\maketitle

\newpage

Recently Ho\v{r}ava proposed a renormalizable gravity theory in four
dimensions which reduces to Einstein gravity with a {\it
non-vanishing} cosmological constant in IR but with improved UV
behaviors \ci{Hora:08,Hora}. Since then various aspects and
solutions have been studied
\ci{Calc,Taka,Kiri,Klus,Lu,Muko:0904,Bran,Cai,Cai:0904,Piao,Gao,Colg,Orla,Keha,Ghod,
Nast,Soti,Muko:0905_1,Nish,Chen:0905_1,Chen:0905_2,Muko:0905_2,Kono,
Char,Li,Kim,Calc:0905}. In \ci{Lu}, it has been pointed out that the
black hole solution in the Ho\v{r}ava model does not recover the
usual Schwarzschild-AdS black hole even though the general
relativity is recovered in IR at the action level. (For a
generalization to topological black holes, see \ci{Cai}.) On the
other hand, in \ci{Keha} an IR modification which allows the flat
Minkowski vacuum has been studied by introducing a term proportional
to the Ricci scalar of the three-geometry $\mu ^4 R^{(3)}$, while
considering ``vanishing" cosmological constant ($\sim \La_W$) in the
Ho\v{r}ava gravity. (For related discussions, see also
\ci{Nast,Soti}.)

In this paper, I consider the black hole and cosmological solutions
in the generalized model with the IR modification term $\mu ^4
R^{(3)}$ but with an ``arbitrary" cosmological constant in the
Ho\v{r}ava gravity. These solutions represent the analogs of the
standard Schwarzschild-(A)dS solutions which have been absent in the
original Ho\v{r}ava model. I discuss their thermodynamical
properties also.

To this ends, I start by considering the ADM decomposition of the
metric
\begin{\eq}
ds^2=-N^2 c^2 dt^2+g_{ij}\left(dx^i+N^i dt\right)\left(dx^j+N^j
dt\right)\
\end{\eq}
and the IR-modified Ho\v{r}ava action which reads
\begin{\eq}
S &= & \int dt d^3 x
\sqrt{g}N\left[\frac{2}{\kappa^2}\left(K_{ij}K^{ij}-\lambda
K^2\right)-\frac{\kappa^2}{2\nu^4}C_{ij}C^{ij}+\frac{\kappa^2
\mu}{2\nu^2}\epsilon^{ijk} R^{(3)}_{i\ell} \nabla_{j}R^{(3)\ell}{}_k
\right.
\nonumber \\
&&\left. -\frac{\kappa^2\mu^2}{8} R^{(3)}_{ij}
R^{(3)ij}+\frac{\kappa^2 \mu^2}{8(3\lambda-1)}
\left(\frac{4\lambda-1}{4}(R^{(3)})^2-\Lambda_W R^{(3)}+3
\Lambda_W^2\right)+\frac{\kappa^2 \mu^2 \om}{8(3\lambda-1)}
R^{(3)}\right]\ , \label{horava}
\end{\eq}
where
\begin{\eq}
 K_{ij}=\frac{1}{2N}\left(\dot{g}_{ij}-\nabla_i
N_j-\nabla_jN_i\right)\
 \end{\eq}
is the extrinsic curvature,
\begin{\eq}
 C^{ij}=\epsilon^{ik\ell}\nabla_k
\left(R^{(3)j}{}_\ell-\frac{1}{4}R^{(3)} \delta^j_\ell\right)\
 \end{\eq}
is the Cotton tensor,  $\kappa,\lambda,\nu,\mu, \La_W$, and $\om$
are constant parameters. The last term, which has been introduced in
\ci{Hora,Keha,Nast}, represents a ``soft" violation of the
``detailed balance" condition in \ci{Hora} and this modifies the IR
behaviors \footnote{In \ci{Keha}, $\omega=8
\mu^2(3\lambda-1)/\kappa^2$ has been considered for the $AdS$ case,
but $\om$ may be considered as an independent parameter, more
generally. }.

Let us consider now a static, spherically symmetric solution with
the metric ansatz
\begin{\eq}
  ds^2=-N(r)^2 c^2 dt^2+\frac{dr^2}{f(r)}+r^2
\left(d\theta^2+\sin^2\theta d\phi^2\right)\ .
\end{\eq}
By substituting the metric ansatz into the action (\ref{horava}),
the resulting reduced Lagrangian, after angular integration, is
given by
\begin{\eq}
{\cal{L}}&=&\frac{\kappa^2\mu^2}{8(1-3\lambda)}\frac{N}{\sqrt{f}}\left[(2\lambda-1)\frac{(f-1)^2}{r^2}
-2\lambda\frac{f-1}{r}f'+\frac{\lambda-1}{2}f'^2\right.\nonumber
\\ &&\left. ~~-2 (\om-\La_W) (1-f-rf') - 3 \La_W^2 r^2 \right]\ ,
\end{\eq}
where
the prime $(')$ denotes the derivative with respect to $r$. In
\ci{Keha}, only the asymptotically Minkowski solution with $\La_W
\ra 0$ limit was considered. Here, I obtain the general solution
with an ``arbitrary" $\La_W$.

The equations of motions are
\begin{\eq}
&&(2\lambda-1)\frac{(f-1)^2}{r^2}-
2\lambda\frac{f-1}{r}f'+\frac{\lambda-1}{2}f'^2-2 (\om-\La_W)
(1-f-rf')- 3 \La_W^2 r^2 =0\ ,
\nonumber \\
&& \left(\frac{N}{\sqrt{f}}\right)' \left((\lambda-1)f'-2\lambda
\frac{f-1}{r}+2(\om-\La_W)
r\right)+(\lambda-1)\frac{N}{\sqrt{f}}\left(f''-\frac{2(f-1)}{r^2}\right)=0
\end{\eq}
by varying the functions $N$ and $f$, respectively.

For the $\lambda=1$
case, which reduces
to the standard Einstein-Hilbert action in the IR limit, I obtain
 \begin{\eq}
N^2=f=1+(\om-\La_W) r^2-\sqrt{r[\omega (\om-2 \La_W) r^3 + \be]}\ ,
\label{solution}
\end{\eq}
 where $\be$ is an integration constant\footnote{If one add another IR
 modification term ${\kappa^2 \mu^2}({8 (3 \la-1)})^{-1} \hat{\be} \La_W^2$ as in
 \ci{Hora,Nast},
 the solution becomes
  $N^2=f=1+(\om-\La_W) r^2-\sqrt{r[\{\omega (\om-2 \La_W)+ \hat{\be} \La_W^2/3 \} r^3
  + \be]}$. But this can be obtained by redefining the parameters
  $\La_W \ra \sqrt{1-\hat{\be}/3}~ \La_W,~ \om \ra \om +
  (\sqrt{1-\hat{\be}/3}-1) \La_W$ in (\ref{solution}).
  }.
It is easy to see that this reduces to L\"{u}, Mei, and Pope (LMP)'s
{\it AdS} black hole solution in \ci{Lu} (I consider $N^2=f$ always,
from now on), by identifying $\be=-\al^2/\La_W$,
\begin{\eq}
f=1-\La_W r^2-\f{\al}{\sqrt{-\La_W}} \sqrt{r}
\end{\eq}
for $\om=0$, Kehagias and Sfetsos's asymptotically flat solution in
\ci{Keha}, by identifying $\be= 4 \om M$,
\begin{\eq}
f=1+\om r^2-\sqrt{r[\omega^2  r^3 + 4 \om M]}
\end{\eq}
for $\La_W=0$.

For $r \gg [\be /\om ( \om-2 \La_W)]^{1/3}$ (by considering
asymptotically AdS case of $\La_W <0$ with $\omega
>0$, $\beta>0$, for the moment) (\ref{solution}) behaves
as
\begin{\eq}
f=1+\f{\La_W^2}{2 \om} r^2 - \f{\be}{2 \sqrt{ \om ( \om-2 \La_W)}}
\f{1}{r} + {\cal O}(r^{-4}). \label{approx}
\end{\eq}
This agrees with the usual \Sch black hole (by adopting the units of
$G=c\equiv 1$)
\begin{\eq}
f=1 - \f{2 M}{r  }
\end{\eq}
for $\La_W=0$ and with $\be=4 \om M$, independently of $\om$. But
for $\La_W \neq 0$, there are corrections in the numerical factors
due to $\om$ effect: With $\be=4 \om M$, (\ref{approx}) can be
re-written as
\begin{\eq}
f=1+\f{|\La_W|}{2}\left|\f{\La_W}{ \om}\right| r^2 - \f{2
M}{\sqrt{1+2 |\La_W/\om|}} \f{1}{r} + {\cal O}(r^{-4})
\label{approx2}
\end{\eq}
in which the coefficients slightly disagree with those of the
standard Schwarzschild-AdS black hole\footnote{This seems to be a
quite generic behavior of the {\it broken} ``detailed balance". See,
for example, \ci{Lu}.}, by the factor `$|\La_W/\om|$',
\begin{\eq}
f=1 +\f{|\La_W|}{2}r^2- \f{2 M}{r}.
\end{\eq}
This solution has a curvature singularity with the power of
$r^{-3/2}$ at $r=0$,
\begin{\eq}
R&=&\f{-15 (\om M)^{1/2}}{2 r^{3/2}} +12 (\om-\La_W) +{\cal
O}(r^{3/2}), \no \\
R^{\mu \nu \al \be}R_{\mu \nu \al \be} &=&\f{81 \om M}{4 r^3} -
\f{30 (\om M)^{1/2} (\om-\La_W)}{r^{3/2}} +{\cal O}(1),
\end{\eq}
but {\it no} curvature singularity at $r=\infty$. Note that this
singularity is {\it milder} than that of Einstein gravity which has
$R^{\mu \nu \al \be}R_{\mu \nu \al \be} \sim r^{-6}$ at
$r=0$.\footnote{I consider the four-dimensional curvature invariants
just for a formal reason, i.e., the comparison with those of
Einstein gravity. But, the degree of singularity is unchanged even
if the three-dimensional curvature invariants are considered only.}

For asymptotically AdS, i.e.,  $\La_W<0$ (with $\om>0$), the
solution (\ref{solution}) has two horizons generally and the
temperature for the outer horizon $r_+$ is given by\footnote{Due to
the lack of Lorentz invariance in UV, the very meaning of the
horizons and Hawking temperature would be changed from the
conventional ones. The light cones would differ for different
wavelengths and so different particles with different dispersion
relations would see different Hawking temperature $T_{H}$ and
entropies, the Hawking spectrum would not be thermal. But from the
recovered Lorentz invariance in IR (with $\la=1$), the usual meaning
of the horizons and $T$ as the Hawking temperature would be
``emerged" for long wavelengths. The calculation and meaning of the
temperature should be understood in this context. }
\begin{\eq}
T=\f{3 \La_W^2 r_+^4 +2 (\om-\La_W)r_+^2 -1 }{ 8 \pi r_+ (1
+(\om-\La_W)r_+^2)}. \label{temp}
\end{\eq}
In Fig.1, the temperature $T$ vs. the horizon radius $r_+$ is
plotted and this shows that \asy, i.e., for large $r_+$, the
temperature interpolates between the AdS cases (above two curves)
and flat (bottom curve). There exists an extremal black hole limit
of the vanishing temperature where the inner horizon $r_-$ meets
with the outer horizon $r_+$ at\footnote{For $\La_W \ra 0$ limit,
(\ref{extrem}) becomes $0/0$. But, from (\ref{temp}), one can get
easily $r_+^*=1/\sqrt{2 \om}$ without the ambiguity.}
\begin{\eq}
r_+^{*} =\sqrt{ \f{-(\om-\La_W)+ \sqrt{ (\om-\La_W)^2 +3 \La_W^2}
}{3 \La_W^2}} \label{extrem}
\end{\eq}
and the integration constant
\begin{\eq}
\be=\f{1+2( \om-\La_W) r_+^2 + \La_W^2 r_+^4 }{r_+}
\end{\eq}
gets the minimum.
\begin{figure}
\includegraphics[width=10cm,keepaspectratio]{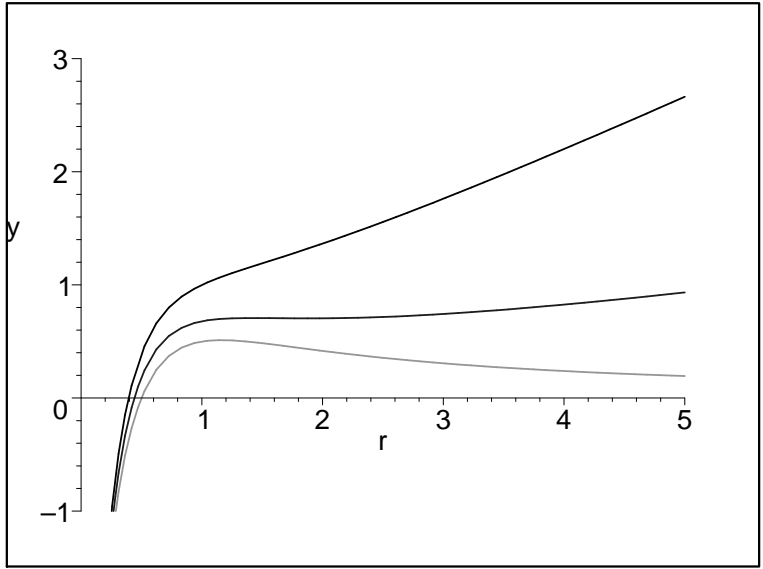}
\caption{Plots of $4 \pi T$ vs. black hole's outer horizon radius
$r_+$ for various $\La_W$, i.e., $\La_W=-1,~ 0.5,~ 0$ (top to
bottom) with $\om=2$. For large black holes, the temperature
interpolates between the AdS cases (above two curves) and flat
(bottom curve). In IR region, i.e., small black holes, there exists
an extremal black hole limit of the vanishing temperature at $r_+^*$
in which $r_-$ meets with the outer horizon $r_+$. The region
smaller than the extremal radius, which being the Cauchy horizon,
shows an unphysical negative temperature.} \label{fig:Nariai}
\end{figure}
The extremal radius $r_+^*$ is the Cauchy horizon and so
continuation to region of $r_+ <r_+^*$ does not make sense to
outside observer; $T<0$ for $r_+<r_+^*$ reflects a pathology of the
region.
(For some recent related discussions, see \ci{Myun}.)

For asymptotically dS, ie., $\La_W >0$, the action is given by an
analytic continuation\footnote{This corresponds to the analytic
continuation of the three-dimensional {\it Euclidian} action
$W_{\rm{Euc}}=\frac{1}{\nu^2}\int Tr \left(\Gamma\wedge
d\Gamma+\frac{2}{3} \Gamma\wedge\Gamma \wedge\Gamma \right)+\mu \int
d^3 x \sqrt{g}(R^{(3)}-2\La_W)$ into $i W_{\rm{Lor}}$ with the
real-valued action $W_{\rm{Lor}}$ \ci{Hora,Lu}. This prescription
agrees with \ci{Park:07} but disagrees with \ci{Witt}. }
\begin{\eq}
\m \ra i \m,~\nu^2 \ra -i \nu^2,~\om \ra -\om
\end{\eq}
of (\ref{horava}) \ci{Lu}.
This can be easily seen in the expansion (\ref{approx2}) for $r \gg
[\be /|\om ( \om-2 \La_W)|]^{1/3}$,
\begin{\eq}
f=1-\f{\La_W}{2}\left|\f{\La_W}{ \om}\right| r^2 - \f{2 M}{\sqrt{1+2
|\La_W/\om|}} \f{1}{r} + {\cal O}(r^{-4})
\end{\eq}
which agrees with the usual Schwarzschild-dS cosmological solution
\begin{\eq}
f=1 -\f{\La_W}{2}r^2- \f{2 M}{r}, \label{Sch-dS}
\end{\eq}
up to some numerical factor corrections. The dS solution also has
two horizons generally; the larger one $r_{++}$ for the cosmological
horizon and the smaller one $r_+$ for the black hole horizon.

The black hole temperature is given by (\ref{temp}) also but now
with $\La_W >0, \om <0$. There exists also an extremal limit of the
vanishing temperature at $r_+^*$ of (\ref{extrem}) in which the
black hole horizon $r_+$ coincides with the cosmological horizon
$r_{++}$, i.e., the {\it Nariai} limit. But a peculiar thing is that
there is an infinite discontinuity of temperature at
\begin{\eq}
\tilde{r}_+=\f{1}{\sqrt{\La_W-\om}}
\end{\eq}
(see Fig. 2).
\begin{figure}
\includegraphics[width=10cm,keepaspectratio]{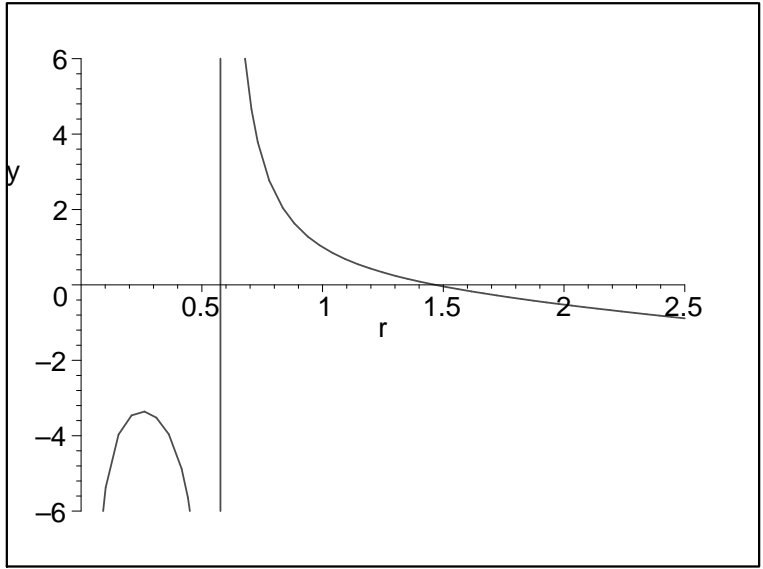}
\caption{Plots of $4 \pi T$ vs. black hole horizon radius $r_+$ for
the dS black hole case with $\La_W=+1$ and $\om=-2$. There exists
also an extremal limit of the vanishing temperature at $r_+^*$ of
(\ref{extrem}) in which the black hole horizon $r_+$ coincides with
the cosmological horizon $r_{++}$, i.e., the {\it Nariai} limit.
There is an infinite discontinuity of temperature at $
\tilde{r}_+={1}/{\sqrt{\La_W-\om}}$.} \label{fig:Nariai}
\end{figure}
This may be understood from the following facts. First, by writing
$\beta=4 \om M$ with mass parameter $M$, in conformity with the
usual convention of (\ref{Sch-dS}), one needs to consider
additionally the condition
\begin{\eq}
M \leq \f{(2 \La_W - \om)}{4} r_+^3 \label{cond}
\end{\eq}
in order that the black hole horizon exists and the curvature
singularity at $r=0$ is not naked. But this inequality is satisfied
always from the relation
\begin{\eq}
M=\f{1+ 2 (\om -\La_W) r_+^2 +\La_W^2 r_+^4}{4 \om r_+}:
\label{mass}
\end{\eq}
(\ref{cond}) reduces to the condition $[(\om -\La_W) r_+^2 +1]^2
\geq 0$, where the equality for $r_+=1/\sqrt{\La_W-\om}=\tilde{r}_+$
corresponds to the upper bound of the mass (\ref{cond}). In other
words, $M < \f{(2 \La_W - \om)}{4} r_+^3$ for all $r_+$ except for
$r_+=\tilde{r}_+$, where $M$ meets the upper bound
\begin{\eq}
M_{\rm{bound}}=\f{(2 \La_W - \om)}{4} r_+^3=\f{2 \La_W - \om}{4 (
\La_W - \om)^{3/2}}.
\end{\eq}
(See Fig.3 for the graphical explanation of this circumstance.)
\begin{figure}
\includegraphics[width=10cm,keepaspectratio]{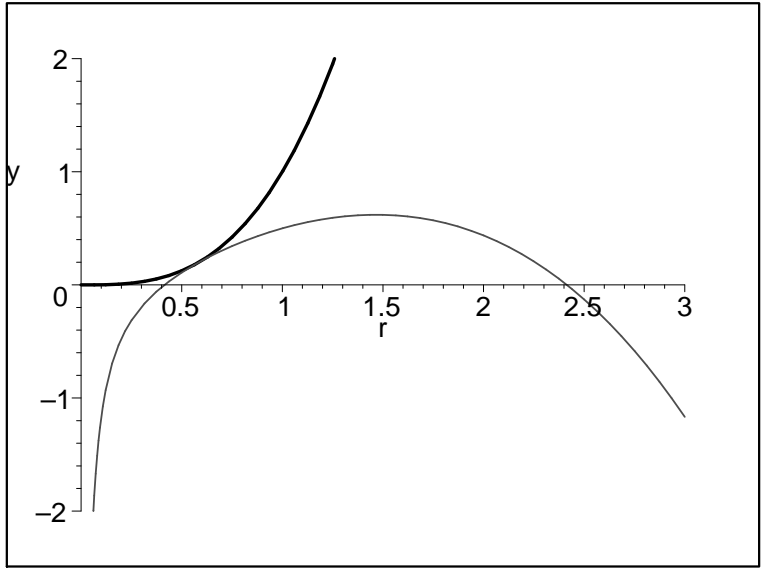}
\caption{Plots of mass spectrum $M$ of (\ref{mass}) (lower curve)
vs. the upper mass bound $M =\f{(2 \La_W - \om)}{4} r_+^3$ of
(\ref{cond}) (upper curve) for the existence of black hole horizon
$r_+$. This shows that the mass bound is always satisfied for all
horizon radius $r_+$ and this is saturated at the point
$r_+={1}/{\sqrt{\La_W-\om}}=\tilde{r}_+$. And also, the mass
parameter gets the maximum value in the spectrum at the Nariai limit
$r_+^*$. Here I plotted $\La_W=+1$ and $\om=-2$ cases but the
results are generally valid for arbitrary values of $\La_W>0$ and
$\om<0$. } \label{fig:Nariai}
\end{figure}
So, the occurrence of the infinite temperature discontinuity and
even the negative temperature for $r_+ < \tilde{r}_+$ would be a
reflection of being the upper bound of the mass parameter $M$, for a
given $r_+$, like the spin system with the upper bound of the energy
level. (For some recent discussions in different contexts, see also
\ci{Park:06}.) There is no ``geometrical" reason to exclude $r_+ <
\tilde{r}_+$, where $T<0$. But, the negative temperature {\it might}
be a signal of the instability of the smaller black hole, like the
negative temperature spin systems in the ordinary surroundings with
a positive temperature. This situation is quite different from the
asymptotically AdS or flat case, where $T<0$ region is geometrically
protected by the Cauchy horizon at $r_+^*$ in which the inner and
outer horizon coincides and $T=0$.

So far, I have studied the black hole and cosmological solutions for
$\la=1$, which matches exactly with the Einstein-Hilbert action in
IR. It would be interesting to find the more general solutions for
arbitrary values of $\la$. 
Especially in cosmology, the arbitrary $\la$ solutions would be also
quite important for the practical purpose \ci{Keha}. So I consider a
homogeneous and isotropic cosmological solution to the action
(\ref{horava}) with the standard FRW form (by recovering ``$c$")
\begin{\eq}
ds^2=-c^2dt^2+a^2(t)\left[\frac{dr^2}{1-kr^2/R_0^2}+r^2\left(d\theta^2+\sin^2\theta
d\phi^2\right)\right],
\end{\eq}
where $k=+1,0,-1$ correspond to a closed, flat, and open universe,
respectively, and $R_0$ is the radius of curvature of the universe
in the current epoch. Assuming the matter contribution to be of the
form of a perfect fluid with the energy density $\rho$ and pressure
$p$, I find that
\begin{\eq}
\left(\f{\dot{a}}{a}\right)^2&=&\frac{\kappa^2}{6(3\lambda-1)}
\left[\rho \pm \frac{3\kappa^2\mu^2}{8(3\lambda-1)} \left( \f{-
k^2}{R_0^4 a^4}+ \f{2 k (\La_W -\om)}{R_0^2 a^2}- \La_W^2 \right) \right] ,  \label{F1}\\
\f{\ddot{a}}{a}&=&\frac{\kappa^2}{6(3\lambda-1)} \left[-\f{1}{2}
(\rho+3 p) \pm \frac{3 \kappa^2\mu^2}{8(3\lambda-1)} \left( \f{
k^2}{R_0^4 a^4}- \La_W^2 \right) \right]. \label{F2}
\end{\eq}
(I have corrected some typos in \ci{Keha}.) Here I have considered
the analytic continuation $\mu^2 \ra - \mu^2$ for the dS case, i.e.,
$\La_W>0$ \ci{Lu} and the upper (lower) sign denotes the AdS (dS)
case.
Note that the $1/a^4$ term, which is the contribution from the
higher-derivative terms in the action (\ref{horava}), exists only
for $k \neq 0$ and become dominant for small $a(t)$, implying that
the cosmological solutions of general relativity are recovered at
large scales. The first Friedman equation (\ref{F1}) generalizes
those of \ci{Lu} and \ci{Keha} to the case with an arbitrary
cosmological constant and the soft IR modification term in
\ci{Hora,Keha,Nast}. However, it is interesting to note that there
is {\it no} contribution from the soft IR modification to the second
Friedman equation (\ref{F2}) and this is identical to that of
\ci{Lu}.

For vacuum solutions with $p=\rho=0$, I have
\begin{\eq}
\left(\f{\dot{a}}{a}\right)^2&=& \mp \frac{\kappa^4
\mu^2}{16(3\lambda-1)^2} \left[ \left( \f{k}{R_0^2 a^2}-\La_W
\right)^2+\f{2 k \om}{R_0^2 a^2}\right].
\end{\eq}
Here, the role of the $\om$ term is crucial. Without that term, only
the constant solution of $a^2=-1/\La_W R_0^2$ with $k=-1$ exists
when $\La_W<0$, otherwise $(\dot{a}/a)^2$ becomes negative \ci{Lu}.
But now with the last term I have more possibilities: There may
exist {\it non-constant} solutions even for $k=+1$ if $-\om$
is big enough to make $(\dot{a}/a)^2>0$. Actually, for $-\om > 2
|\La_W|,~k=+1$, there exists a cyclic universe solution
\begin{\eq}
a^2_{\rm AdS}(t)= \f{k (-\om- |\La_W|)}{R_0^2 \La^2_W} \left[ 1 \pm
\sqrt{\f{ -\om (-\om-2 |\La_W|)}{(-\om -|\La_W|)^2}} \left|\rm{sin}
\left(\f{\kappa^2 \mu \La_W}{2 (3 \lambda-1)} (t-\gamma) \right)
\right|\right]
\end{\eq}
which is oscillating between the inner and outer bouncing scale
factors
\begin{\eq}
a^{\pm}_{\rm AdS} = \f{\sqrt{-2 k \om} \pm \sqrt{2 k ( -\om-2 |
\La_W|)}}{2 R_0 |\La_W |}
\end{\eq}
and the integration constant $\gamma$, depending on the initial
conditions. The two bouncing scale factors merge as $-\om$ becomes
smaller and coincide at $a^{\pm}_{\rm AdS}= \sqrt{k /|\La_W| R_0^2
}$ when $-\om= 2 |\La_W |$. For $k=-1$, the solution reduces to
LMP's constant solution when $\om=0$, but there is no solution for
other values when $\om <0$.

 On the other hand, for the dS case, i.e., $\La_W>0$ and $\om >0$,
the general solution is given by
\begin{\eq}
a^2_{\rm dS}(t)=\f{ 2 |3 \lambda-1| }{\kappa^2 |\mu| R_0^2 \La_W}
{\rm e}^{\pm \f{\kappa^2 |\mu|   \La_W}{2 |3 \lambda-1|}
(t-\gamma)}+\f{k^2 \kappa^2 |\mu| \om (\om-2 \La_W)}{ 8 |3
\lambda-1| R_0^2 \La_W} {\rm e}^{\mp \f{\kappa^2 |\mu| \La_W}{2 |3
\lambda-1|} (t-\gamma)} -\f{ k(\om- \La_W)}{R_0^2 \La_W^2}.
\end{\eq}
For $k=-1$, a bounce occurs at
\begin{\eq}
a^{\pm}_{\rm dS} = \f{\sqrt{-2 k \om} \pm \sqrt{-2 k ( \om-2
\La_W)}}{2 R_0 \La_W }
\end{\eq}
when $\om > 2 \La_W$; at $a^+_{\rm dS}$ when $a(t)$ shrinks toward
$a^+_{\rm dS}$, at $a^-_{\rm dS}$ when $a(t)$ expands toward
$a^-_{\rm dS}$. $\om = 2 \La_W$ is the marginal case where the two
bouncing scale factors coincide at $a^{\pm}_{\rm dS}= \sqrt{- k
/\La_W R_0^2}$ and the universe evolves monotonically from that
point to de Sitter vacuum asymptotically or vice versa. When $\om <
2 \La_W$ (and also for arbitrary values of $\om >0$ when $k=+1$),
the universe evolves from the big bang singularity to de Sitter
vacuum or vice versa. For $\om=0,~k=+1$, this reduces to the LMP's
solution with the minimum scale factor $a_{\rm{min}}=1/\sqrt{\La_W}
R_0$ \ci{Lu}\footnote{It is interesting to note that these various
scenarios for dS case have been considered earlier by Calcagni in
the original Ho\v{r}ava model \ci{Calc} but it is important to note
that these can be ``realized" only in our modified model with the
$\om$ terms. In Calcagni's notation, the scenarios are categorized
by the values of $3 c^2 \tilde{K}^2 - 4 B^2 |\tilde{\La}|$
but one can have only $3 c^2 \tilde{K}^2 - 4 B^2 |\tilde{\La}|=0$ if
one use the actual values of the parameters for the Ho\v{r}ava
model, i.e., (34) and (35). This corresponds to the LMP's
solution.}.

In addition to the evolution of the universe, there are very strong
constraints on the equation of state parameters for the constituents
of our universe. So, it would be an important test of our Ho\v{r}ava
gravity, whose additional contributions to the Friedman equation may
not be distinguishable from those of dark energy, to see if one can
meet the correct observational constraints\footnote{While this paper
was being finalized, I met a paper by Mukohyama \ci{Muko:0905_2}
which propose the dark ``matter" as integration constant in
Ho\v{r}ava gravity.}. Then, it is easy to see that the energy
density and pressure of the dark energy part are given by (for a
related discussion with matters in the context of the original
Ho\v{r}ava gravity, see \ci{Sari})
\begin{\eq}
\rho_{\rm D.E.}&=&\pm \frac{3\kappa^2\mu^2}{8(3\lambda-1)} \left(
\f{-
k^2}{R_0^4 a^4}- \f{2 k \om}{R_0^2 a^2}- \La_W^2 \right), \\
p_{\rm D.E.}&=&\mp \frac{\kappa^2\mu^2}{8(3\lambda-1)} \left( \f{
k^2}{R_0^4 a^4}- \f{2 k \om}{R_0^2 a^2}-3 \La_W^2 \right),
\end{\eq}
respectively and the equation of state parameter is given by
\begin{\eq}
w_{\rm D.E.}=\f{p_{\rm D.E.}}{\rho_{\rm D.E.}}=\left( \f{k^2- 2 k
\om R_0^2 a^2-3 \La_W^2  R_0^4 a^4}{3k^2+ 6 k \om R_0^2 a^2+3
\La_W^2 R_0^4 a^4} \right).
\end{\eq}
This interpolates from $w_{\rm D.E.}=1/3$ in the UV limit to $w_{\rm
D.E.}=-1$ in the IR limit but the detailed evolution patten in
between them depends on the parameters $k, \om, \La_W$ (See
Fig.4-7). This looks to be consistent with current observational
constraints but it seems to be still too early to decide what the
right one is \ci{Ries}. But if I consider the transition point from
deceleration phase to acceleration phase, which is given by
$a_T=\sqrt{|k|/|\Lambda_W| R_0^2}$ from (\ref{F2}) by neglecting the
matter contributions, the formula gives $w_{\rm D.E.}=-1/3$, {\it
independently} of the parameters $k, \om, \La_W$. If I use $a_T \sim
1/1.03 \approx 0.9709 $ which corresponds to $z_T\sim 0.30$ in the
astronomer's parametrization $z=1/a-1$, I get $|\Lambda_W|\sim
(1.03)^2 R_0^{-2} \approx 1.0609 R_0^{-2}$ for the non-flat universe
with $|k|=1$. And also, if I use $\Omega_k \sim -0.026$ \ci{Sper} in
the current epoch $(a=1)$ for the deviation from the critical
density, $\Omega_k=\mu^2 k |\Lambda_W| L_{\rm P}^2/2a^2 H^2 R_0^2
M_{\rm P}^2$, Hubble parameter $H\equiv \dot{a} /a$, the ratio of
Planck mass and length $M_{\rm P}/L_{\rm P} \equiv 2 (3 \lambda
-1)/\kappa^2$, $k=-1$, I get $\mu \sim 0.2214 H_0 R_0 M_{\rm
P}/L_{\rm P}$ with the current value of Hubble parameter
$H_0$.\footnote{I follow the physical convention of Ryden \ci{Ryde}
which disagrees with \ci{Hora:08,Hora}.} Finally, if I use $w_{\rm
D.E.}\sim -1.08$ in the current epoch, I get $\om \sim 1.0067
R_0^{-2}$ which predicts the evolution of $w_{\rm D.E.}$ as one of
the curves in Fig.6 since $\om <|\Lambda_W|$: If I use $R_0 \sim
6.2017~ c/H_0$ from $\Om_k=k c^2/H_0^2 R_0^2 \sim -0.026$ and $H_0
\sim 70 {\rm km~ s^{-1} Mpc^{-1}}$, I get $\Lambda_W \sim 1.5018
\times 10^{-9} \rm{Mpc}^{-2}, ~\om \sim 1.4251 \times 10^{-9}
\rm{Mpc}^{-2},~\mu\sim 5.6636 \times 10^{35} \rm{kg~ s^{-1}}$.

Note added: After finishing this paper, a related paper \ci{Wang}
appeared whose classification of all the possible cosmology
solutions in the Ho\v{r}ava gravity {\it without} the detailed
balance
is overlapping with mine. (See also \cite{Mina}.)

\begin{figure}
\includegraphics[width=10cm,keepaspectratio]{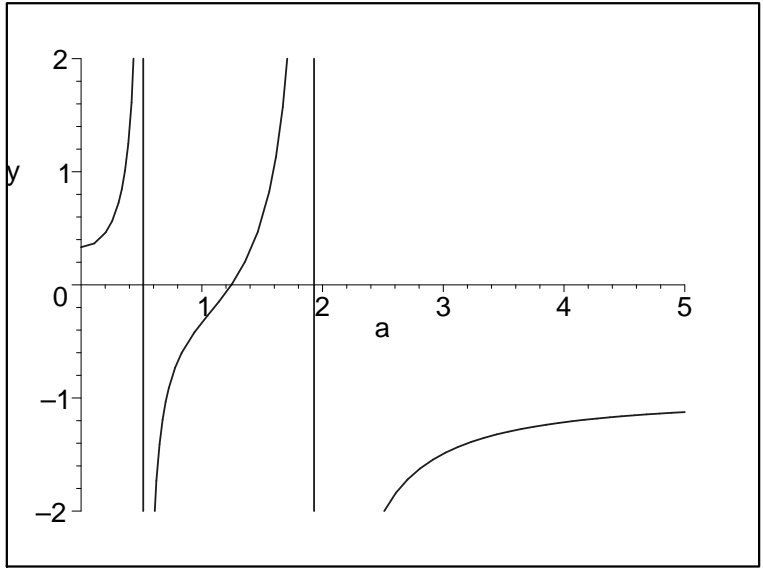}
\caption{Plot of equation of state parameter $w_{\rm D.E.}$ vs.
scale factor $a(t)$ for $\om^2 > \La_W^2,~k \om<0$. There are two
infinite discontinuities of $w_{\rm D.E.}$ at $\tilde{a}^{\pm} = {
\sqrt{ -k \om \pm |k| \sqrt{\om^2-\La_W^2} }}/{ |\La_W| R_0 }$ where
$\rho_{\rm D.E.}$ vanishes. Here, I considered $|\om| R_0^2 =2,
|\Lambda_W| R_0^2 =1$
case ($\om R_0^2 =-2,
k=+1$ or
$\om R_0^2 =+2,
k=-1)$.}\label{fig:EOS_Singular1}
\end{figure}

\begin{figure}
\includegraphics[width=10cm,keepaspectratio]{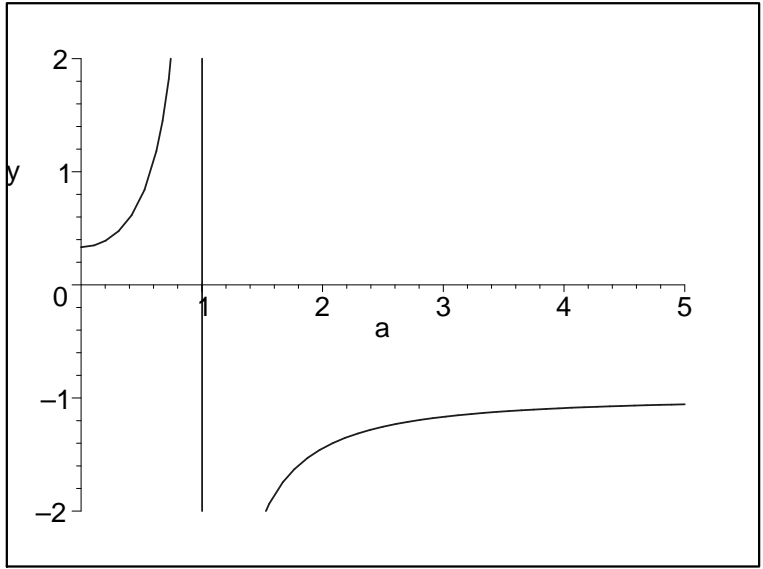}
\caption{Plot of equation of state parameter $w_{\rm D.E.}$ vs.
scale factor $a(t)$ for $\om^2 = \La_W^2,~k \om <0$.  The two points
of infinite discontinuities $\tilde{a}^{\pm}$ in Fig.4 merge as
$|\om|$ approaches to $|\La_W|$ and they meet at $\tilde{a}^{\pm} =
{ \sqrt{ |k| }}/{ |\La_W| R_0 }$ when $\om^2 = \La_W^2$. In this
plot, I considered $|\om| R_0^2 =|\Lambda_W| R_0^2 =1$ ($\om R_0^2
=-1,
k=+1$
or $\om R_0^2 =+1,
k=-1$
). }
\label{fig:EOS_Singular2}
\end{figure}

\begin{figure}
\includegraphics[width=10cm,keepaspectratio]{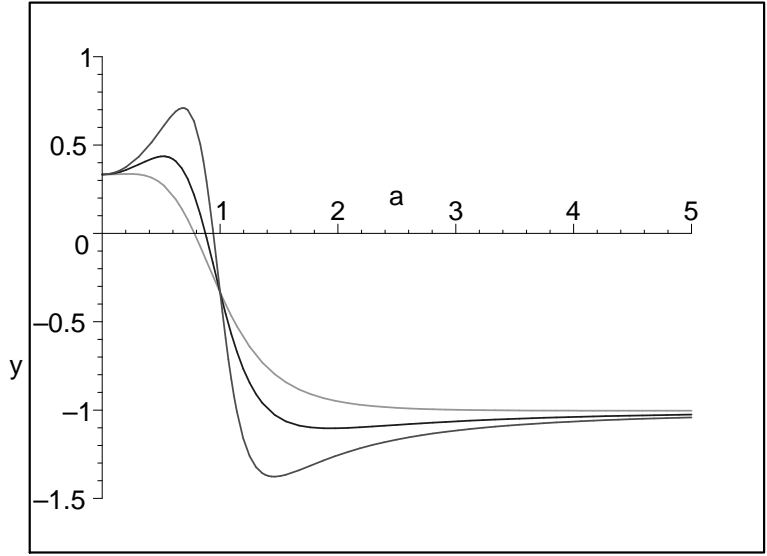}
\caption{Plots of equation of state parameters $w_{\rm D.E.}$ vs.
scale factor $a(t)$ for $\om^2 < \La_W^2,~k \om<0$ ( $\om R_0^2
=+1/1.3, +1/2, +1/10$, $k=-1$
or $\om R_0^2 =-1/1.3, -1/2, -1/10$
, $k=+1$ with $|\La_W| R_0^2 =1$ (top to bottom in the left region )
). When $|\om|$ is not far from $|\La_W|$, there is a region where
$w_{\rm D.E.}$ is fluctuating beyond the UV and IR limits and this
can be understood as a smooth deformation of the plot of Fig.5. When
$|\om|$ is small enough, $w_{\rm D.E.}$ is monotonically decreasing
from $1/3$ in the UV limit to $-1$ in the IR limit.  }
\label{fig:EOS_open_Regular}
\end{figure}

\begin{figure}
\includegraphics[width=10cm,keepaspectratio]{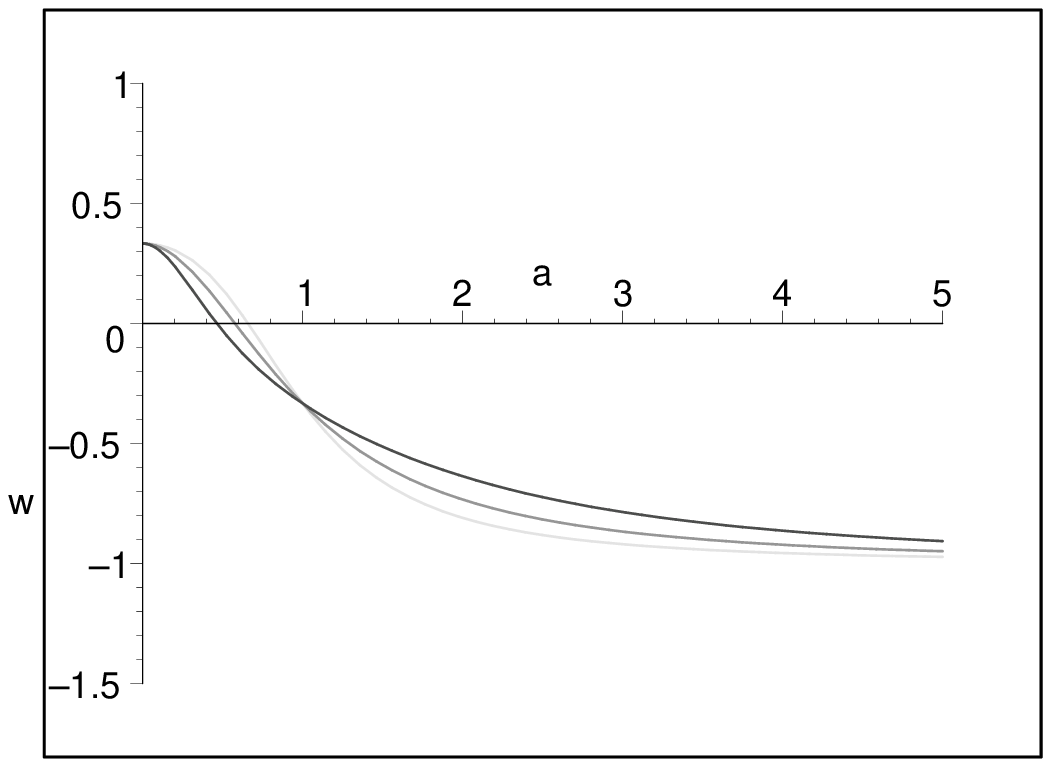}
\caption{Plots of equation of state parameters $w_{\rm D.E.}$ vs.
scale factor $a(t)$ for $k \om>0$
( $\om R_0^2 =+2, +1, +1/2$, $k=+1$ or $\om R_0^2 =-2, -1, -1/2$,
$k=-1$ with $|\La_W| R_0^2 =1$ (top to bottom in the left region ).
In this case, $w_{\rm D.E.}$ is ``always" monotonically decreasing
from $1/3$ in the UV limit to $-1$ in the IR limit.   }
\label{fig:EOS_closed_Regular}
\end{figure}

\section*{Acknowledgments}

I would like to thank Hyeong-Chan Kim, Gungwon Kang, Jungjai Lee for
inviting me to San-Gol Mauel Workshop and all other attendants for
discussing and drawing my attention to Ho\v{r}ava gravity and
especially for Hyunseok Yang and Inyong Cho for reviewing the
subject. And also I would like to thank the organizers of APCTP-BLTP
JINR Workshop ``Frontiers in Black Hole Physics at Dubna" where I
have finished this work. This work was supported by the Korea
Research Foundation Grant funded by Korea Government(MOEHRD)
(KRF-2007-359-C00011).

\newcommand{\J}[4]{#1 {\bf #2} #3 (#4)}
\newcommand{\andJ}[3]{{\bf #1} (#2) #3}
\newcommand{\AP}{Ann. Phys. (N.Y.)}
\newcommand{\MPL}{Mod. Phys. Lett.}
\newcommand{\NP}{Nucl. Phys.}
\newcommand{\PL}{Phys. Lett.}
\newcommand{\PR}{Phys. Rev. D}
\newcommand{\PRL}{Phys. Rev. Lett.}
\newcommand{\PTP}{Prog. Theor. Phys.}
\newcommand{\hep}[1]{ hep-th/{#1}}
\newcommand{\hepp}[1]{ hep-ph/{#1}}
\newcommand{\hepg}[1]{ gr-qc/{#1}}
\newcommand{\bi}{ \bibitem}

\end{document}